\documentclass[sigconf]{acmart}

\usepackage{booktabs}
\usepackage{xspace}
\usepackage{amssymb}
\usepackage{listings}
\usepackage{tikz}
\usepackage{amsmath}
\usepackage{algorithm}
\usepackage[noend]{algpseudocode}
\usepackage{pifont}
\usepackage{tabularx,booktabs}
\usepackage{flushend}
\usepackage{amsmath}
\usepackage{caption}
\usepackage{subcaption}
\PassOptionsToPackage{hyphens}{url}

\usepackage{enumitem}
\usepackage[T1]{fontenc}
\usepackage{hyperref}
\usepackage{multirow}
\usepackage{pgfplots}
\usepackage{soul}
\usepackage{boxedminipage}
\usepackage{hyperref}

\definecolor{darkred}{rgb}{0.75,0,0}
\definecolor{darkblue}{rgb}{0,0,0.75}
\definecolor{darkgreen}{rgb}{0,0.75,0}

\newcommand{\hide}[1]{}

\definecolor{pblue}{rgb}{0.13,0.13,1}
\definecolor{pgreen}{rgb}{0,0.5,0}
\definecolor{pred}{rgb}{0.9,0,0}
\definecolor{pgrey}{rgb}{0.46,0.45,0.48}
\algnewcommand\algorithmicswitch{\textbf{switch}}
\algnewcommand\algorithmiccase{\textbf{case}}
\algnewcommand\algorithmicassert{\texttt{assert}}
\algnewcommand\Assert[1]{\State \algorithmicassert(#1)}%
\algdef{SE}[SWITCH]{Switch}{EndSwitch}[1]{\algorithmicswitch\ #1\ \algorithmicdo}{\algorithmicend\ \algorithmicswitch}%
\algdef{SE}[CASE]{Case}{EndCase}[1]{\algorithmiccase\ #1}{\algorithmicend\ \algorithmiccase}%
\algtext*{EndSwitch}%
\algtext*{EndCase}%

\usetikzlibrary{fit,calc,trees,positioning,arrows,chains,shapes.geometric,decorations.pathreplacing,decorations.pathmorphing,shapes,matrix,shapes.symbols}
\tikzset{
>=stealth',
  punktchain/.style={
    rectangle, 
    rounded corners, 
    draw=black, very thick,
    text width=5em, 
    minimum height=3em, 
    text centered, 
    on chain},
  line/.style={draw, thick, <-},
  element/.style={
    tape,
    top color=white,
    bottom color=blue!50!black!60!,
    minimum width=5em,
    draw=blue!40!black!90, very thick,
    text width=5em, 
    minimum height=3.5em, 
    text centered, 
    on chain},
  every join/.style={->, thick,shorten >=1pt},
  tuborg/.style={decorate},
  tubnode/.style={midway, right=2pt},
}

\newcounter{myobjctr}
\newenvironment{objective}%
{
\noindent\refstepcounter{myobjctr}\begin{boxedminipage}{\linewidth}\textbf{Challenge \themyobjctr:}
}{
\end{boxedminipage}
\smallskip
}

\newcommand*\circled[1]{\raisebox{1pt}{\scalebox{.7}{\tikz[baseline=(char.base)]{
            \node[shape=circle,draw,inner sep=2pt] (char) {#1};}}}}

\setcopyright{none}
\renewcommand\footnotetextcopyrightpermission[1]{}
\begin{document}

\title{Self-adaptive static analysis}
\author{Eric Bodden}
\affiliation{Paderborn University \& Fraunhofer IEM}
\email{eric.bodden@upb.de}

\begin{abstract}
Static code analysis is a powerful approach to detect quality deficiencies such as performance bottlenecks, safety violations or security vulnerabilities already during a software system's implementation. Yet, as current software systems continue to grow, current static-analysis systems more frequently face the problem of insufficient scalability. We argue that this is mainly due to the fact that current static analyses are implemented fully manually, often in general-purpose programming languages such as Java or C, or in declarative languages such as Datalog. This design choice predefines the way in which the static analysis evaluates, and limits the optimizations and extensions static-analysis designers can apply.

To boost scalability to a new level, we propose to fuse static-analysis with just-in-time-optimization technology, introducing for the first time static analyses that are managed and inherently self-adaptive. Those analyses automatically adapt themselves to yield a performance/precision tradeoff that is optimal with respect to the analyzed software system and to the analysis itself. 

Self-adaptivity is enabled by the novel idea of designing a dedicated intermediate representation, not for the analyzed program but for the analysis itself. This representation allows for an automatic optimization and adaptation of the analysis code, both ahead-of-time (through static analysis of the static analysis) as well as just-in-time during the analysis' execution, similar to just-in-time compilers.
\end{abstract}



\keywords{Static program analysis, virtual machines, intermediate representations}

\maketitle

\section{Introduction}
\label{sec:intro}
Over the past decades, static code analysis has made significant progress, and has seen many novel applications: originally used mainly for the purpose of
ahead-of-time program optimization~\cite{Sundaresan:2000:PVM:353171.353189,vallee2000optimizing}, it has now become also a common tool for program understanding
as well as for finding software quality defects, in particular security vulnerabilities~\cite{ltoimpact,yan2014leakchecker,FlowDroid}. This success is due to
decades of static-analysis research, which yielded the discovery of novel algorithms, data structures and design principles that make static analyses more
precise and scalable than ever before.~\cite{ase15,oopsla17ideal,sui2012static,xiao2011geometric,Smaragdakis:2011:PYC:1926385.1926390}

Yet at the same time, the size of software systems has grown immensely. Simple smartphones now run applications as large and complex as the most complex
server-side applications just ten years ago. Hence, while the progress in static-analysis research is significant in absolute terms, one must fear that
nevertheless the technology will always lack behind the software applications' increase in size and complexity.
To break this barrier, one requires nothing short of a breakthrough in static-analysis technology.

Such a break-through is currently hindered by the fact that, so far, all known static-analysis tools have been implemented by hand, and in general-purpose programming languages such as Java or C/C++, or in some cases partly in Datalog. Most static analyses require only a limited expressiveness, and thus often can be expressed as pushdown-problems~\cite{reps2005weighted} or even graph-reachability
problems~\cite{Reps:1995:PID:199448.199462}, certainly most often do not require a Turing-complete language. Since optimizations for pushdown automata and graph algorithms are well studied, one would think it possible to just apply powerful automated optimizations to such analyses. Yet, the current state of the art is to implement static analyses themselves in general-purpose programming languages. In our view, those languages are too expressive: they are, in fact, Turing complete, which greatly hinders powerful automated optimizations of the static analyses. 

In this work we propose a novel fusion of static-analysis and just-in-time-optimization technology, yielding static analyses that are inherently self-adaptive, and use this self-adaptivity for self-optimization.  Current analyses are not self-adaptive because their evaluation strategy is, at least for the most part, hard-coded. While analysis implementations might select among a set of multiple pre-defined evaluation strategies depending on the analysis problem and analyzed application at hand, the possible choices and the selection strategies themselves are fixed. Moreover, once a strategy has been selected, it is executed by instantiating pre-defined static-analysis components.

We instead envision a solution that produces for each concrete analysis problem a highly customized and optimized static-analysis implementation, specifically tailored to the problem at hand. Moreover, the solution should have the ability to re-adapt and thus further optimize this implementation based on an introspection into the analysis' own execution. This is what we mean when we speak of self-adaptivity, and this is where lessons learned from research on just-in-time optimization will be useful. Enabling such self-adaptivity requires one to design and implement the analysis according to a completely novel engineering methodology, a description of which is the core contribution of this paper.

Developing a working system fully implementing the idea we propose is a multi-person-year effort. In this paper we restrict ourselves to explaining the core idea and to posing the main research challenges one needs to address to obtain a working solution. That way we hope that the software engineering research community will join us in our quest for an optimal solution strategy.  

Section~\ref{sec:concept} presents the core concepts and challenges of our proposal, while Section~\ref{sec:related} situates the proposal into related work. Section~\ref{sec:conclusion} concludes.


\section{Core Concepts and Challenges}
\label{sec:concept}
\begin{figure*}[btp]
\centering
\includegraphics[width=\textwidth]{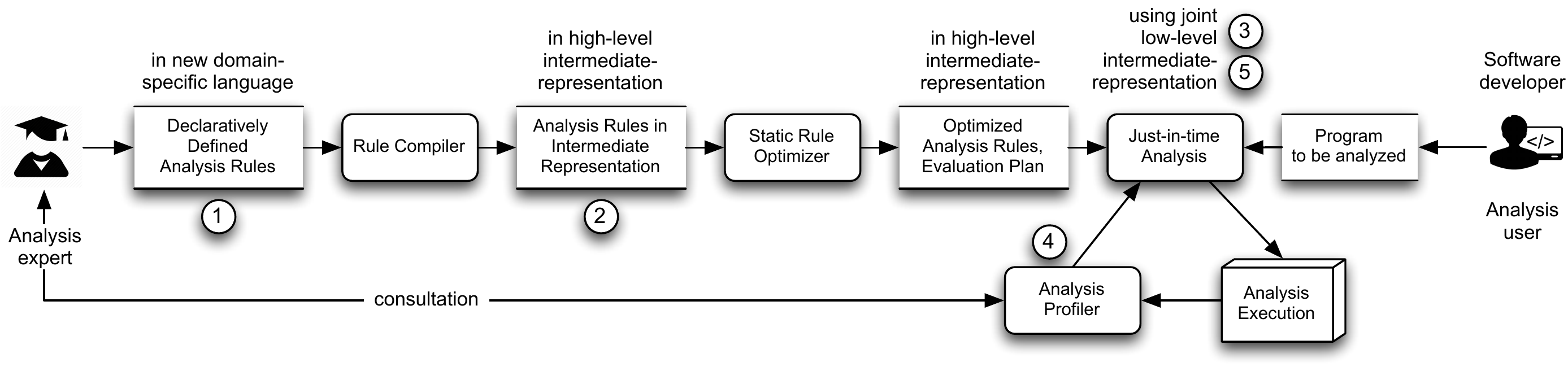}
  \caption{Workflow of the envisioned self-adaptive static analysis}
\label{fig:flowchart}
\end{figure*}

The main objective of this work is to enable self-adaptive, self-optimizing static analyses. This, in turn, requires one to design and implement the analysis according to a completely novel engineering methodology, yielding an architecture outlined in Figure~\ref{fig:flowchart}. The architecture fuses a number of concepts known from the area of static program analysis with those of just-in-time program optimizations, as we know it from application-level virtual machines. Importantly, though, the concepts have been fused such that not the program under analysis is the one that's being optimized but instead \emph{the just-in-time-optimization targets the static analysis itself!}

In the following we outline the various core concepts and objectives that Figure~\ref{fig:flowchart} presents. 

\paragraph{\protect\circled{1}~Declarative definition language for static analysis}
At the core of our proposal is the following paradigm shift: to represent the program analysis itself not in a general-purpose programming language
such as Java or C/C++, (nor a general-purpose logic programming language such as Datalog) but rather in a domain-specific language that is optimally amenable to domain-specific optimizations, i.e., optimizations that are only correct when taking into account specific knowledge about the domain of static program analyses. The challenge in designing such a language is that it must have exactly the right level of expressiveness: it must be expressive enough to cover a reasonably broad set of possible static-analysis clients, yet at the same time its expressiveness must be restricted enough such as to allow for powerful automated domain-specific optimizations. Moreover, the language must be easy enough for static-analysis users, i.e., typically for software developers, to write and understand, yet at the same time must be machine-readable as well. Using the language, it should be possible to conveniently express also large sets of static-analysis rules, as is common, for instance in security code analysis tools, which commonly check programs against hundreds of vulnerability patterns.

\begin{objective}\label{obj:dsl}
The design and implementation of a domain-specific language (DSL) for expressing static-analysis rules.
\end{objective}

\paragraph{\protect\circled{2}~High-level intermediate representation}
While the DSL above must be easy for humans to understand, to enable automated optimizations of the static analysis we instead envision a dedicated domain-specific intermediate representation (IR) of the static analysis, which is not designed to be optimally understandable by humans, but instead to be optimally amenable to domain-specific optimizations, i.e., optimizations that are only correct when taking
into account specific knowledge about the domain of static program analyses. The focus of the proposed high-level intermediate representation is on optimizing static-analyses \emph{ahead of time}, i.e., prior to their execution. Similarly to query optimizations in database research, such an IR will allow one to exploit synergies between similar analysis rules, and to find an optimal evaluation strategy---at this point independent of the analyzed program.

\begin{objective}\label{obj:hlir}
The design, and implementation of a high-level intermediate analysis representation, with the goal to allow ahead-of-time analysis optimizations.
\end{objective} 

\paragraph{\protect\circled{3}~Low-level intermediate representation}
The above ahead-of-time optimizations have the goal to improve analysis performance (while
maintaining precision) \emph{without} taking the analyzed program into account. Yet, previous experience shows that the analyzed program can have a great
influence on what is the optimal configurations for a static analysis of that program. \cite{Smaragdakis:2011:PYC:1926385.1926390,bodden2007staged} We hence desire a mechanism to allow domain-specific optimizations that take program characteristics into account, allowing the analysis configuration to be optimally tuned to the analyzed program while the analysis is being conducted. This requires one to alter the analysis execution just-in-time, or at least from one execution to the next. While in theory this might be able by altering its high-level intermediate representation (Challenge~\ref{obj:hlir}), it is likely that one can benefit from an additional low-level representation that is less declarative but closer to the analysis' actual execution. This also gives one the opportunity to combine in the same representation not just aspects of the analysis but also aspects of the program to be analyzed. Opposed to the high-level IR, to allow for just-in-time adaptation, for the low-level IR it must be possible to alter the analysis representation in place, potentially re-generating analysis code in the fly, similarly to how current virtual machines re-generate program code at runtime. 

\begin{objective}\label{obj:llir}
The design, and implementation of a low-level intermediate representation for just-in-time optimizations during analysis execution, combining both aspects of the  analysis and the program to be analyzed.
\end{objective} 

\paragraph{\protect\circled{4}~Static-analysis profiler}
A self-optimizing static analyzer, but also a human static-analysis expert, will require deep insights into the analysis' own execution, and in particular its performance hotspots. The fact that the analysis is self-adaptive actually makes this harder than normal, as the analysis code that actually executes is not code hand-written by the analysis designer, but instead code generated from the analysis' intermediate representation. Such a self-adaptive design thus threatens to lose the link between the analysis definition and its execution. To address this challenge, we identify the objective of designing, implementing and evaluating a dedicated profiling tool for self-adaptive static analyses. The profiling tool is meant to automatically identify execution hotspots that cause the analysis performance to degrade, for instance regions of the analyzed program that cause the analysis to iterate for exceedingly long times, or cause it to consume unusually large amounts or memory. The profiler should moreover link back those performance hotspots to the relevant fragments of the static analysis responsible for those parts of the analysis execution, in the different IRs as well as in the original analysis definition in our novel DSL. The profiler also directly links back to our JIT analysis engine, which we describe next.

\begin{objective}\label{obj:prof}
The design and implementation of a profiling tool for static-analysis runs, providing traceability into the different analysis representations.
\end{objective} 

\paragraph{\protect\circled{5}~Automated just-in-time optimization}
The final step to towards our main objective is to develop an optimization engine that uses domain knowledge about the static analysis it executes, paired with information from the profiled analysis run, to determine more optimal analysis execution strategies, and to trigger those strategies through an automated self-adaptation. We envision this engine to effectively implement a control loop, in which profiling information is continuously processed to determine the optimality of the current strategy, and to determine other strategies that might be more efficient (or which may save more memory) during the remainder of the execution and/or during the next analysis run. One challenge in this space is to find the right machine learning algorithms to create a reasonable situational awareness of the executing analysis, and to draw the right conclusions from the observed profiles. Another challenge lies in the analysis adaptations. Some simple adaptations are implemented easily. For instance one can change a taint analysis from forward to backward if encountering in a program significantly more sources than sinks, thus yielding an analysis runs with a comparatively low number of taints. Or else, when computing a constant propagation, if observing that parts of a program conduct linear arithmetic only, one can use an efficient IFDS-based~\cite{Reps:1995:PID:199448.199462} tabulation solver to solve a linear-constant propagation problem for those program parts. Some adaptations, however, might involve more low-level adaptations of the executing analysis. For instance, one might think of actually implementing parts of the analysis using certain data structures, either tuned for runtime or memory efficiency, depending on where more efficiency is required. To this end, the optimization engine must be able to closely interact with the low-level intermediate representation (Challenge~\ref{obj:llir}).

\begin{objective}\label{obj:mlopt}
An optimization engine that learns from profiled analysis runs in order to better optimize future runs or even the remainder of a current run. 
\end{objective} 

\section{State of the Art and Related Work}
\label{sec:related}
\label{sec:sota}

We next describe how related work from the state-of-the-art literature can fuel the research we propose here. Space restrictions force us to only discuss the
most related areas of research.

\paragraph*{Application-level virtual machines} \label{sec:vms}
In terms of its architecture, the solution we propose here has a strong similarity with the architecture of application-level virtual machines such as the Java virtual machine (JVM). Their just-in-time compilers (JIT) profile the hosted applications' execution to optimize that very same execution on the fly. Some virtual machines further include ahead-of-time optimizations and pool optimized code for efficient reuse on later runs.\cite{haggar2005single} And also here one can observe a tradeoff between automation and understandability: debugging and optimizing virtual machines is notoriously hard, which is why researchers have developed dedicated tools for this purpose.\cite{wimmer2013maxine} Those tools typically visualize the executing program's code in the different intermediate representations on which the JVM itself operates. The main difference between a JVM and its JIT to our proposal is that here we are not executing general-purpose Java code but instead a specific static analysis. Moreover, we seek to have domain-specific representations of this static analysis at all levels of its execution. Thus, while one might, in fact, reuse ideas regarding the architecture and design of application-level virtual machines, the specific intermediate representations, the transformations between them, and the optimizations within them will greatly differ from those of existing approaches. 

\paragraph*{Declarative languages used for static analysis}
Not all static analysis are implemented in Turing-complete general-purpose languages. The static-analysis framework Doop~\cite{bravenboer2009strictly}, for instance, implements its static analyses in the logic programming language Datalog. Using Datalog gives users the great advantage that the implemented static-analysis rules are relatively simple to write, read and reason about. The declarative nature of the language also means that---in theory---users need to not be concerned with how the rules are evaluated, as this decision is entirely left to the Datalog engine. Yet, as past experience with Doop and similar approaches has shown, to make Datalog-based analyses truly efficient, one still requires optimizations on two levels.

First, automated optimizations on the level of the Datalog language itself. Doop is frequently used in combination with highly optimized Datalog solvers such as LogicBlox~\cite{aref2015design} or Souffle\cite{jordan2016souffle}. The latter is a Datalog solver specifically designed for the purpose of supporting static analyses: it translates the datalog rules into C code implementing a highly optimized Datalog solver for the particular rule set at hand. To some extent Souffle is thus similar to what we propose here, but the main difference of Souffle and all other Datalog engines, compared to what we propose here, is that Datalog, albeit being a logic programming language, is still a general-purpose language. It is not domain-specific, and thus has no domain-specific constructs that would make the static-analysis solution particularly efficient to compute. In result, while users benefit from the declarative nature of the language, they only benefit from automated general-purpose Datalog optimizations, which are limited because they lack domain knowledge. A particular limitation of Datalog, namely the lack of being able to express fixed-point iterations, which are commonplace in static analysis, triggered the incarnation of the novel logic programming language Flix~\cite{madsen2016datalog}. Flix is essentially an extension of Datalog with certain primitives that allows analysis writers to directly express fixed-point computation. Yet, the tool that currently implements Flix is, again, nothing more but a general-purpose solver for the Flix language. It is not a static-analysis tool, and does not support any kind of further domain-specific optimizations.


\paragraph*{Conceptual analysis frameworks}
Some domain-specific optimizations that a just-in-time optimizer should consider for static analysis can be obtained by choosing the optimal \emph{analysis framework}. By analysis framework we here mean a framework in the conceptual sense. For instance, two major conceptual frameworks are known to compute correct solutions for context-sensitive inter-procedural analysis problems: the so-called \emph{call-strings approach} and the \emph{functional approach}~\cite{sharir1978two}. The call-strings approach has the advantage that it can be applied to pretty much any static data-flow analysis, specifically any such analysis that fits the monotone framework~\cite{kam1977monotone}. Yet, it has the drawback that it could analyze callee procedures while distinguishing more contexts than necessary, thus unnecessarily wasting computation time. Moreover, one must bound the amount of context the analysis uses, and choosing the bound poorly might jeopardize performance, precision and even correctness~\cite{khedker2009data}. The functional approach has the advantage of providing unlimited context-sensitivity (hence no bound is required), yet is only tractable for analysis problems whose merge operator is set union and whose flow functions distribute over this merge operator---so-called IFDS or IDE problems\cite{Reps:1995:PID:199448.199462}. In cases where the analysis problem fits such a framework, one has the advantage that one can compute precise solutions (equivalent to the theoretically optimal but generally uncomputable MOP solution~\cite{khedker2009data}), and moreover that highly efficient solution algorithms exist. Moreover, in recent work, we were able to show that it is sometimes possible to decompose such analysis problems that are actually not distributive, for instance pointer analysis or general constant propagation, into sub-problems that are, in fact, distributive. Thus they can be efficiently solved using IFDS or IDE solvers, assuming some extra analysis code that then composes the results of individual distributive computations to the final analysis result. In our experience, such an approach showed significant speedups and gains in precision when compared to previous approaches. One goal of the optimization engine we seek to develop here will be to identify the potential for such a decomposition of analysis problems automatically.

\section{Conclusion}
\label{sec:conclusion}
We have presented the novel idea of enabling self-adaptive, self-optimizing static analyses, by developing a dedicated domain-specific language and multiple
dedicated domain-specific intermediate representations (IR) not to express programs but to express the analysis itself. A novel low-level IR, in
particular, is meant to represent how a particular analysis executes on a particular program. This enables just-in-time optimizations making use of both
static-analysis and program properties. We have presented an overall solution architecture that has some resemblence to application-level just-in-time
optimizers, and have highlighted the core challenges the community will face in developing a concrete solution.


\bibliographystyle{ACM-Reference-Format}
\bibliography{references}


\begin{thebibliography}{00}


\ifx \showCODEN    \undefined \def \showCODEN     #1{\unskip}     \fi
\ifx \showDOI      \undefined \def \showDOI       #1{#1}\fi
\ifx \showISBNx    \undefined \def \showISBNx     #1{\unskip}     \fi
\ifx \showISBNxiii \undefined \def \showISBNxiii  #1{\unskip}     \fi
\ifx \showISSN     \undefined \def \showISSN      #1{\unskip}     \fi
\ifx \showLCCN     \undefined \def \showLCCN      #1{\unskip}     \fi
\ifx \shownote     \undefined \def \shownote      #1{#1}          \fi
\ifx \showarticletitle \undefined \def \showarticletitle #1{#1}   \fi
\ifx \showURL      \undefined \def \showURL       {\relax}        \fi
\providecommand\bibfield[2]{#2}
\providecommand\bibinfo[2]{#2}
\providecommand\natexlab[1]{#1}
\providecommand\showeprint[2][]{arXiv:#2}

\bibitem[\protect\citeauthoryear{Aref, ten Cate, Green, Kimelfeld, Olteanu,
  Pasalic, Veldhuizen, and Washburn}{Aref et~al\mbox{.}}{2015}]%
        {aref2015design}
\bibfield{author}{\bibinfo{person}{Molham Aref}, \bibinfo{person}{Balder ten
  Cate}, \bibinfo{person}{Todd~J Green}, \bibinfo{person}{Benny Kimelfeld},
  \bibinfo{person}{Dan Olteanu}, \bibinfo{person}{Emir Pasalic},
  \bibinfo{person}{Todd~L Veldhuizen}, {and} \bibinfo{person}{Geoffrey
  Washburn}.} \bibinfo{year}{2015}\natexlab{}.
\newblock \showarticletitle{Design and implementation of the LogicBlox system}.
  In \bibinfo{booktitle}{{\em Proceedings of the 2015 ACM SIGMOD International
  Conference on Management of Data}}. ACM, \bibinfo{pages}{1371--1382}.
\newblock


\bibitem[\protect\citeauthoryear{Arzt, Rasthofer, Fritz, Bodden, Bartel, Klein,
  Le~Traon, Octeau, and McDaniel}{Arzt et~al\mbox{.}}{2014}]%
        {FlowDroid}
\bibfield{author}{\bibinfo{person}{Steven Arzt}, \bibinfo{person}{Siegfried
  Rasthofer}, \bibinfo{person}{Christian Fritz}, \bibinfo{person}{Eric Bodden},
  \bibinfo{person}{Alexandre Bartel}, \bibinfo{person}{Jacques Klein},
  \bibinfo{person}{Yves Le~Traon}, \bibinfo{person}{Damien Octeau}, {and}
  \bibinfo{person}{Patrick McDaniel}.} \bibinfo{year}{2014}\natexlab{}.
\newblock \showarticletitle{{FlowDroid}: Precise Context, Flow, Field,
  Object-sensitive and Lifecycle-aware Taint Analysis for Android Apps}. In
  \bibinfo{booktitle}{{\em PLDI 2014}}. \bibinfo{publisher}{ACM},
  \bibinfo{address}{New York, NY, USA}, \bibinfo{pages}{259--269}.
\newblock
\showISBNx{978-1-4503-2784-8}
\showDOI{%
\url{https://doi.org/10.1145/2594291.2594299}}


\bibitem[\protect\citeauthoryear{Bodden, Hendren, and Lhot{\'a}k}{Bodden
  et~al\mbox{.}}{2007}]%
        {bodden2007staged}
\bibfield{author}{\bibinfo{person}{Eric Bodden}, \bibinfo{person}{Laurie
  Hendren}, {and} \bibinfo{person}{Ondrej Lhot{\'a}k}.}
  \bibinfo{year}{2007}\natexlab{}.
\newblock \showarticletitle{A staged static program analysis to improve the
  performance of runtime monitoring}. In \bibinfo{booktitle}{{\em ECOOP 2007}}.
  Springer-Verlag, \bibinfo{pages}{525--549}.
\newblock


\bibitem[\protect\citeauthoryear{Bravenboer and Smaragdakis}{Bravenboer and
  Smaragdakis}{2009}]%
        {bravenboer2009strictly}
\bibfield{author}{\bibinfo{person}{Martin Bravenboer} {and}
  \bibinfo{person}{Yannis Smaragdakis}.} \bibinfo{year}{2009}\natexlab{}.
\newblock \showarticletitle{Strictly declarative specification of sophisticated
  points-to analyses}.
\newblock \bibinfo{journal}{{\em ACM SIGPLAN Notices\/}} \bibinfo{volume}{44},
  \bibinfo{number}{10} (\bibinfo{year}{2009}), \bibinfo{pages}{243--262}.
\newblock


\bibitem[\protect\citeauthoryear{Haggar, Mickelson, and Wendt}{Haggar
  et~al\mbox{.}}{2005}]%
        {haggar2005single}
\bibfield{author}{\bibinfo{person}{Peter~F Haggar}, \bibinfo{person}{James~A
  Mickelson}, {and} \bibinfo{person}{David Wendt}.}
  \bibinfo{year}{2005}\natexlab{}.
\newblock \bibinfo{title}{Single-instance class objects across multiple JVM
  processes in a real-time system}.
\newblock   (\bibinfo{date}{Jan.~11} \bibinfo{year}{2005}).
\newblock
\newblock
\shownote{US Patent 6,842,759.}


\bibitem[\protect\citeauthoryear{Jordan, Scholz, and Suboti{\'c}}{Jordan
  et~al\mbox{.}}{2016}]%
        {jordan2016souffle}
\bibfield{author}{\bibinfo{person}{Herbert Jordan}, \bibinfo{person}{Bernhard
  Scholz}, {and} \bibinfo{person}{Pavle Suboti{\'c}}.}
  \bibinfo{year}{2016}\natexlab{}.
\newblock \showarticletitle{Souffl{\'e}: On synthesis of program analyzers}. In
  \bibinfo{booktitle}{{\em CAV 2016}}. Springer, \bibinfo{pages}{422--430}.
\newblock


\bibitem[\protect\citeauthoryear{Kam and Ullman}{Kam and Ullman}{1977}]%
        {kam1977monotone}
\bibfield{author}{\bibinfo{person}{John~B Kam} {and} \bibinfo{person}{Jeffrey~D
  Ullman}.} \bibinfo{year}{1977}\natexlab{}.
\newblock \showarticletitle{Monotone data flow analysis frameworks}.
\newblock \bibinfo{journal}{{\em Acta Informatica\/}} \bibinfo{volume}{7},
  \bibinfo{number}{3} (\bibinfo{year}{1977}), \bibinfo{pages}{305--317}.
\newblock


\bibitem[\protect\citeauthoryear{Khedker, Sanyal, and Sathe}{Khedker
  et~al\mbox{.}}{2009}]%
        {khedker2009data}
\bibfield{author}{\bibinfo{person}{Uday Khedker}, \bibinfo{person}{Amitabha
  Sanyal}, {and} \bibinfo{person}{Bageshri Sathe}.}
  \bibinfo{year}{2009}\natexlab{}.
\newblock \bibinfo{booktitle}{{\em Data flow analysis: theory and practice}}.
\newblock \bibinfo{publisher}{CRC Press}.
\newblock


\bibitem[\protect\citeauthoryear{Lerch, Sp\"ath, Bodden, and Mezini}{Lerch
  et~al\mbox{.}}{2015}]%
        {ase15}
\bibfield{author}{\bibinfo{person}{Johannes Lerch}, \bibinfo{person}{Johannes
  Sp\"ath}, \bibinfo{person}{Eric Bodden}, {and} \bibinfo{person}{Mira
  Mezini}.} \bibinfo{year}{2015}\natexlab{}.
\newblock \showarticletitle{Access-Path Abstraction: Scaling Field-Sensitive
  Data-Flow Analysis With Unbounded Access Paths}. In \bibinfo{booktitle}{{\em
  ASE 2015}}. \bibinfo{pages}{619--629}.
\newblock
\showURL{%
\url{http://www.bodden.de/pubs/lsb+15access-path.pdf}}


\bibitem[\protect\citeauthoryear{Lillack, K\"astner, and Bodden}{Lillack
  et~al\mbox{.}}{2014}]%
        {ltoimpact}
\bibfield{author}{\bibinfo{person}{Max Lillack}, \bibinfo{person}{Christian
  K\"astner}, {and} \bibinfo{person}{Eric Bodden}.}
  \bibinfo{year}{2014}\natexlab{}.
\newblock \showarticletitle{Tracking Load-time Configuration Options}. In
  \bibinfo{booktitle}{{\em ASE 2014}}. \bibinfo{pages}{445--456}.
\newblock
\showDOI{%
\url{https://doi.org/10.1145/2642937.2643001}}


\bibitem[\protect\citeauthoryear{Madsen, Yee, and Lhot{\'a}k}{Madsen
  et~al\mbox{.}}{2016}]%
        {madsen2016datalog}
\bibfield{author}{\bibinfo{person}{Magnus Madsen}, \bibinfo{person}{Ming-Ho
  Yee}, {and} \bibinfo{person}{Ond{\v{r}}ej Lhot{\'a}k}.}
  \bibinfo{year}{2016}\natexlab{}.
\newblock \showarticletitle{From datalog to flix: A declarative language for
  fixed points on lattices}. In \bibinfo{booktitle}{{\em PLDI 2016}},
  Vol.~\bibinfo{volume}{51}. ACM, \bibinfo{pages}{194--208}.
\newblock


\bibitem[\protect\citeauthoryear{Reps, Horwitz, and Sagiv}{Reps
  et~al\mbox{.}}{1995}]%
        {Reps:1995:PID:199448.199462}
\bibfield{author}{\bibinfo{person}{Thomas Reps}, \bibinfo{person}{Susan
  Horwitz}, {and} \bibinfo{person}{Mooly Sagiv}.}
  \bibinfo{year}{1995}\natexlab{}.
\newblock \showarticletitle{Precise Interprocedural Dataflow Analysis via Graph
  Reachability}. In \bibinfo{booktitle}{{\em POPL 1995}} {\em
  (\bibinfo{series}{POPL '95})}. \bibinfo{publisher}{ACM},
  \bibinfo{address}{New York, NY, USA}, \bibinfo{pages}{49--61}.
\newblock
\showISBNx{0-89791-692-1}
\showDOI{%
\url{https://doi.org/10.1145/199448.199462}}


\bibitem[\protect\citeauthoryear{Reps, Schwoon, Jha, and Melski}{Reps
  et~al\mbox{.}}{2005}]%
        {reps2005weighted}
\bibfield{author}{\bibinfo{person}{Thomas Reps}, \bibinfo{person}{Stefan
  Schwoon}, \bibinfo{person}{Somesh Jha}, {and} \bibinfo{person}{David
  Melski}.} \bibinfo{year}{2005}\natexlab{}.
\newblock \showarticletitle{Weighted pushdown systems and their application to
  interprocedural dataflow analysis}.
\newblock \bibinfo{journal}{{\em Science of Computer Programming\/}}
  \bibinfo{volume}{58}, \bibinfo{number}{1-2} (\bibinfo{year}{2005}),
  \bibinfo{pages}{206--263}.
\newblock


\bibitem[\protect\citeauthoryear{Sharir and Pnueli}{Sharir and Pnueli}{1978}]%
        {sharir1978two}
\bibfield{author}{\bibinfo{person}{Micha Sharir} {and} \bibinfo{person}{Amir
  Pnueli}.} \bibinfo{year}{1978}\natexlab{}.
\newblock \bibinfo{title}{Two approaches to interprocedural data flow
  analysis}.
\newblock   (\bibinfo{year}{1978}).
\newblock


\bibitem[\protect\citeauthoryear{Smaragdakis, Bravenboer, and
  Lhot\'{a}k}{Smaragdakis et~al\mbox{.}}{2011}]%
        {Smaragdakis:2011:PYC:1926385.1926390}
\bibfield{author}{\bibinfo{person}{Yannis Smaragdakis}, \bibinfo{person}{Martin
  Bravenboer}, {and} \bibinfo{person}{Ondrej Lhot\'{a}k}.}
  \bibinfo{year}{2011}\natexlab{}.
\newblock \showarticletitle{Pick Your Contexts Well: Understanding
  Object-sensitivity}. In \bibinfo{booktitle}{{\em POPL 2011}}.
  \bibinfo{publisher}{ACM}, \bibinfo{address}{New York, NY, USA},
  \bibinfo{pages}{17--30}.
\newblock
\showISBNx{978-1-4503-0490-0}
\showDOI{%
\url{https://doi.org/10.1145/1926385.1926390}}


\bibitem[\protect\citeauthoryear{Sp\"ath, Ali, and Bodden}{Sp\"ath
  et~al\mbox{.}}{2017}]%
        {oopsla17ideal}
\bibfield{author}{\bibinfo{person}{Johannes Sp\"ath}, \bibinfo{person}{Karim
  Ali}, {and} \bibinfo{person}{Eric Bodden}.} \bibinfo{year}{2017}\natexlab{}.
\newblock \showarticletitle{IDEal: Efficient and Precise Alias-aware Dataflow
  Analysis}. In \bibinfo{booktitle}{{\em OOPSLA/SPLASH 2017}}.
  \bibinfo{publisher}{ACM Press}.
\newblock
\newblock
\shownote{To appear.}


\bibitem[\protect\citeauthoryear{Sui, Ye, and Xue}{Sui et~al\mbox{.}}{2012}]%
        {sui2012static}
\bibfield{author}{\bibinfo{person}{Yulei Sui}, \bibinfo{person}{Ding Ye}, {and}
  \bibinfo{person}{Jingling Xue}.} \bibinfo{year}{2012}\natexlab{}.
\newblock \showarticletitle{Static memory leak detection using full-sparse
  value-flow analysis}. In \bibinfo{booktitle}{{\em ISSTA 2012}}. ACM,
  \bibinfo{pages}{254--264}.
\newblock


\bibitem[\protect\citeauthoryear{Sundaresan, Hendren, Razafimahefa,
  Vall{\'e}e-Rai, Lam, Gagnon, and Godin}{Sundaresan et~al\mbox{.}}{2000}]%
        {Sundaresan:2000:PVM:353171.353189}
\bibfield{author}{\bibinfo{person}{Vijay Sundaresan}, \bibinfo{person}{Laurie
  Hendren}, \bibinfo{person}{Chrislain Razafimahefa}, \bibinfo{person}{Raja
  Vall{\'e}e-Rai}, \bibinfo{person}{Patrick Lam}, \bibinfo{person}{Etienne
  Gagnon}, {and} \bibinfo{person}{Charles Godin}.}
  \bibinfo{year}{2000}\natexlab{}.
\newblock \showarticletitle{Practical Virtual Method Call Resolution for Java}.
  In \bibinfo{booktitle}{{\em OOPSLA 2000}} {\em (\bibinfo{series}{OOPSLA
  '00})}. \bibinfo{publisher}{ACM}, \bibinfo{address}{New York, NY, USA},
  \bibinfo{pages}{264--280}.
\newblock
\showISBNx{1-58113-200-X}
\showDOI{%
\url{https://doi.org/10.1145/353171.353189}}


\bibitem[\protect\citeauthoryear{Vall{\'e}e-Rai, Gagnon, Hendren, Lam,
  Pominville, and Sundaresan}{Vall{\'e}e-Rai et~al\mbox{.}}{2000}]%
        {vallee2000optimizing}
\bibfield{author}{\bibinfo{person}{Raja Vall{\'e}e-Rai},
  \bibinfo{person}{Etienne Gagnon}, \bibinfo{person}{Laurie Hendren},
  \bibinfo{person}{Patrick Lam}, \bibinfo{person}{Patrice Pominville}, {and}
  \bibinfo{person}{Vijay Sundaresan}.} \bibinfo{year}{2000}\natexlab{}.
\newblock \showarticletitle{Optimizing Java bytecode using the Soot framework:
  Is it feasible?}. In \bibinfo{booktitle}{{\em Compiler Construction (CC)}}.
  \bibinfo{publisher}{Springer}.
\newblock


\bibitem[\protect\citeauthoryear{Wimmer, Haupt, Van De~Vanter, Jordan,
  Dayn{\`e}s, and Simon}{Wimmer et~al\mbox{.}}{2013}]%
        {wimmer2013maxine}
\bibfield{author}{\bibinfo{person}{Christian Wimmer}, \bibinfo{person}{Michael
  Haupt}, \bibinfo{person}{Michael~L Van De~Vanter}, \bibinfo{person}{Mick
  Jordan}, \bibinfo{person}{Laurent Dayn{\`e}s}, {and} \bibinfo{person}{Douglas
  Simon}.} \bibinfo{year}{2013}\natexlab{}.
\newblock \showarticletitle{Maxine: An approachable virtual machine for, and
  in, java}.
\newblock \bibinfo{journal}{{\em ACM Transactions on Architecture and Code
  Optimization (TACO)\/}} \bibinfo{volume}{9}, \bibinfo{number}{4}
  (\bibinfo{year}{2013}), \bibinfo{pages}{30}.
\newblock


\bibitem[\protect\citeauthoryear{Xiao and Zhang}{Xiao and Zhang}{2011}]%
        {xiao2011geometric}
\bibfield{author}{\bibinfo{person}{Xiao Xiao} {and} \bibinfo{person}{Charles
  Zhang}.} \bibinfo{year}{2011}\natexlab{}.
\newblock \showarticletitle{Geometric encoding: forging the high performance
  context sensitive points-to analysis for Java}. In \bibinfo{booktitle}{{\em
  ISSTA 2011}}. ACM, \bibinfo{pages}{188--198}.
\newblock


\bibitem[\protect\citeauthoryear{Yan, Xu, Yang, and Rountev}{Yan
  et~al\mbox{.}}{2014}]%
        {yan2014leakchecker}
\bibfield{author}{\bibinfo{person}{Dacong Yan}, \bibinfo{person}{Guoqing Xu},
  \bibinfo{person}{Shengqian Yang}, {and} \bibinfo{person}{Atanas Rountev}.}
  \bibinfo{year}{2014}\natexlab{}.
\newblock \showarticletitle{LeakChecker: Practical static memory leak detection
  for managed languages}. In \bibinfo{booktitle}{{\em Proceedings of Annual
  IEEE/ACM International Symposium on Code Generation and Optimization}}. ACM,
  \bibinfo{pages}{87}.
\newblock


\end{thebibliography}


\end{document}